\documentclass[12pt]{iopart}

\usepackage{amssymb}
\usepackage{graphicx}
\usepackage{bm}
\usepackage{color}
\usepackage{cancel}
\usepackage{slashed}
\usepackage{subfigure}

\newcommand{\Ignore}[1]{}

\newcommand{\Ket}[1]{\vert #1\rangle}
\newcommand{\Bra}[1]{\langle #1\vert}

\newcommand{\ii}{\mathrm{i}}
\newcommand{\ee}{\mathrm{e}}

\begin{document}

\title{Sensitivity of Measurement-Based Purification Processes to Inner Interactions}

\author{Benedetto Militello}
\address{Dipartimento di Fisica e Chimica, Universit\`a degli Studi di Palermo, Via Archirafi 36, I-90123 Palermo, Italy}
\address{I.N.F.N. Sezione di Catania}


\author{Anna Napoli}
\address{Dipartimento di Fisica e Chimica, Universit\`a degli Studi di Palermo, Via Archirafi 36, I-90123 Palermo, Italy}
\address{I.N.F.N. Sezione di Catania}

\begin{abstract}
The sensitivity of a repeated measurement-based purification
scheme to additional undesired couplings is analyzed, focusing on
the very simple and archetypical system consisting of two
two-level systems interacting with a repeatedly measured one.
Several regimes are considered and in the strong coupling (i.e.,
when the coupling constant of the undesired interaction is very
large) the occurrence of a quantum Zeno effect is proven to
dramatically jeopardize the efficiency of the purification
process.
\end{abstract}

\maketitle

\section{Introduction}\label{sec:introduction}

Preparation of quantum systems is a basic preliminary step in many protocols and therefore is of fundamental importance in nanotechnology applications, in the field of quantum information~\cite{ref:qinfo}, quantum teleportation~\cite{ref:qtele} and even in quantum thermodynamics~\cite{ref:qthermo}. If the system is in a mixed state, and we want to prepare it into a pure state, no unitary evolution is helpful. The simplest way to obtain a pure state from a non pure one is to perform a measurement on the system in order to exploit the wave function collapse. Nevertheless, more advanced techniques based on measurements exist, one of which consists in performing Quantum Non-Demolition measurements (QND)~\cite{ref:QND1,ref:QND2,ref:QND3,ref:QND4,ref:QND5}. This scheme is based on the idea that the system that we want to
initialize (we will address it main system) is coupled to an auxiliary system through an Hamiltonian that commutes with the
free Hamiltonian of the system (in order to avoid back action), and in addition the auxiliary system is repeatedly measured.
Alternatively, one can relax the condition of commutation between the free system Hamiltonian and the interaction between the main and the auxiliary system, following the scheme in Ref.~\cite{ref:NakazatoPRL2003}. According to such approach, the
system to be prepared is coupled to a repeatedly measured one, without requiring commutation of the interaction with free
Hamiltonians. The net result of this process is the \lq extraction\rq\, of pure states from the initial condition of the
main system. Starting from this scheme the possibility of distilling entangled states between distant systems, independently from the initial conditions, has been brought to light~\cite{ref:CompagnoPRA2004}. This procedure has been also explored
from the theoretical point of view in several directions: the influence of quantum noise during the extraction process has been analyzed in detail~\cite{ref:MilitelloPRA2007,ref:NakazatoPRA2008,ref:Militello2016}, non monotonic behaviors of the purity of the state of the system during the extraction process have been brought to light~\cite{ref:MilitelloPRA2008-bis}. Moreover, the possibility of extracting interesting superpositions of angular momentum eigenstates for two oscillators through repeated measurements on a two-level system has been theoretically proven in the field of trapped ions~\cite{ref:MilitelloPRA2004}. The possibility of totally controlling a qubit state (its purity, energy, etc) through repeated measurments on an ancilla system has also been theoretically demonstrated~\cite{ref:Paternostro2005}. Very
recently, generation of long-lived singlet pairs in a nuclear spin ensemble coupled to the electron spins of a Nitrogen Vacancy center in diamond has been proposed~\cite{ref:Greiner2017}. Over the years, generalizations of the repeated-measurement based purification scheme have been proposed: we mention schemes involving iterative operations (not necessarily
measurments)~\cite{ref:Burgarth2009}  and a two-step measurements scheme~\cite{ref:Qiu2012}. It is the case to underline that very recently many papers on purification protocols, in different physical contexts, have appeared in literature witnessing the importance and the actuality of this topic~\cite{ref:Bernad2016,ref:Grimmer2017,ref:Wallnofer2017,ref:Zhao2017}.

Though the purification scheme of Ref.~\cite{ref:NakazatoPRL2003} is based on repeated measurements at regular time intervals, thus recalling the pattern of the quantum Zeno effect (QZE), the time interval between two measurements is typically not that small as requested for QZE~\cite{ref:Mishra77,ref:PascazioFacchi1,ref:PascazioFacchi2}.

In Ref.~\cite{ref:NakazatoPRA2004} the possibility of extracting entangled states has been investigated in detail, especially considering spin/qubit systems where a repeatedly measured one can induce entanglement between two initially uncorrelated qubits which do not directly interact. Under special conditions (time distance between two measurements and specific results from measurements), the common interaction with a third qubit which is repeatedly measured drives the system made of the other two qubits toward a maximally entangled state. One could expect that switching on a direct interaction between such qubits could improve the probability of inducing entanglement between them, hence making the entanglement extraction process more efficient. Surprisingly, the response of the system is not that simple and the effectiveness of the extraction of entanglement is not necessarily improved by the additional coupling. In fact, in the weak coupling some areas of the parameter space exhibits improvements, some other areas corresponds to diminished stability or efficiency (higher degree of entanglement in the extracted state, higher efficiency of the extraction process), while other areas (usually the bigger part) are more more or less insensitive to the additional coupling. We also analyze the strong coupling regime, where the undesired interaction becomes dominant to the interaction with the repeated measured subsystem. In this regime, the occurrence of a generalized quantum Zeno effect is shown to be responsible for a significant diminishing of the efficiency of the purification process. In both cases, weak and strong coupling limit, we also observe an interesting dependence of the complex phase of the coupling constant of the additional coupling.

The paper is organized as follows. In the next section the repeated measurement-based purification scheme of Ref.~\cite{ref:NakazatoPRL2003} is summarized. In section \ref{sec:model} we introduce the very simple model of two interacting qubits (inner interaction) which in turn interact with a repeatedly measured one (external interaction). This model includes as a particular case the one analyzed in Ref.~\cite{ref:NakazatoPRA2004}, where the inner interaction is absent. In section \ref{sec:stability} we analyze the stability of the purification scheme with respect to the inner interaction, and in section \ref{sec:qze} we give some analytical explanation of some behaviours numerically predicted. Finally, in section \ref{sec:conclusions} we give some conclusive remarks.

\section{Framework}\label{sec:framework}

Here we summarize the purification scheme introduced in Ref.~\cite{ref:NakazatoPRL2003}. Consider a system $\mathrm{S}$ that we want to initialize and an auxiliary system $\mathrm{X}$ that is interacting with the former, repeatedly measured at regular time intervals and always found in the same state, say $\Ket{\psi}_\mathrm{X}$. The effective non unitary dynamics of system $\mathrm{S}$ is well described by the following operator:
\begin{equation}
V(\tau)= _\mathrm{X}\Bra{\psi} \ee^{-\ii H \tau} \Ket{\psi}_\mathrm{X} \,,%
\end{equation}
where $H$ is the total Hamiltonian governing the dynamics of $\mathrm{S}+\mathrm{X}$ and $\tau$ is the time interval between
two measurements. After $N$ steps, the effective dynamics is given by $V(\tau)^N$. Let us denote by $\lambda_1,..., \lambda_M$ the eigenvalues of $V(\tau)$ (ordered in such a way that $|\lambda_i| \ge |\lambda_{i+1}|$), by $\Ket{\lambda_1},..., \Ket{\lambda_M}$ the right eigenvectors and by $\Bra{\tilde{\lambda}_1},..., \Bra{\tilde{\lambda}_M}$ the left eigenvectors, such that $\langle \tilde{\lambda}_i|\lambda_j\rangle=\delta_{ij}$, so that $V(\tau)^N =  \sum_k \lambda_k^N \Ket{\lambda_k}\Bra{\tilde{\lambda}_k}$. If the system $\mathrm{S}$ is initially in the state $\rho_0$, after $N$ steps
it will be in the state
\begin{eqnarray}
  \nonumber
  \rho(N\tau) &=& V(\tau)^N \rho_0 [V^\dag(\tau)]^N \\
  &=& \sum_{kj}
  \Bra{\tilde{\lambda}_k}\rho_0\Ket{\tilde{\lambda}_j}(\lambda_k\lambda_j^*)^{N}
  \Ket{\lambda_k}\Bra{\lambda_j}\,,
\end{eqnarray}
which, under the hypothesis $|\lambda_1|>|\lambda_2|$ and for large enough $N$, will be approximated by the following non normalized state:
\begin{equation}
  \rho(N\tau) =
  \Bra{\tilde{\lambda}_1}\rho_0\Ket{\tilde{\lambda}_1}|\lambda_1|^{2N}
  \Ket{\lambda_1}\Bra{\lambda_1}\,,
\end{equation}
where the lack of normalization expresses the fact that the procedure is a conditional one. Therefore we can reformulate this
result by saying that the system will be fund in the state $\Ket{\lambda_1}$ with a probability
\begin{equation}
  {\cal P}(N,\tau) =
  \Bra{\tilde{\lambda}_1}\rho_0\Ket{\tilde{\lambda}_1}|\lambda_1|^{2N}\,.
\end{equation}

The success of this procedure depends on three factors: (i) the state $\Ket{\lambda_1}$ must be an interesting state (this depends on the specific needs we have); (ii) the number of steps required to extract $\Ket{\lambda_1}$ must be not too large (this depends on the ratio $|\lambda_2/\lambda_1|$: the smaller this ratio the faster the process); (iii) the probability of success should not go to zero, which is related to the fact that $|\lambda_1|^{2N}$ must be non vanishing. This last condition is realized either through the realization of condition (ii) or by fulfilling the condition $|\lambda_1|\approx 1$. In particular, the case $|\lambda_1|=1$ is said case of optimal extraction and exhibits stability with respect to the number of steps, being $|\lambda_1|^{2N}=1$, $\forall N$. Nevertheless, it is important to note that, if the amount of entanglement in the extracted state says us that the we can obtain entanglement and a higher stability allows to obtain a non vanishing  success probability for the purification process, it is the efficiency parameter that determines the very possibility of extracting something. In fact, even in the presence of high entanglement and optimal stability, a low efficiency means that the process will last a very long time, and if the second eigenvalues $\lambda_2$ has the same modulus of $\lambda_1$ then one lose the possibility of extracting anything. In other words, the very possibility (even with low probability) of extracting something (whether an entangled state or not) is given by a non vanishing efficiency.

\section{The Model}\label{sec:model}

Here we consider a three interacting-qubit system, one of which is repeatedly measured in order to purify the state of the other two, and extract entangled states. Moreover, we introduce suitable quantities (witnesses) to quantify the degree of extracted entanglement, the efficiency and the stability of the extraction process.

\subsection{Hamiltonian}

In Ref.\cite{ref:NakazatoPRA2004} it has been investigated the possibility of extracting entanglement between two systems, say $\mathrm{A}$ and $\mathrm{B}$ through the interaction with a third system $\mathrm{X}$ which is repeatedly measured at regular time intervals and always found in the same state $\Ket{\psi}_\mathrm{X}$.  The Hamiltonian of such three-qubit system is the following:
\begin{eqnarray}
  H = \sum_{k=\mathrm{A},\mathrm{B},\mathrm{X}} \frac{\omega}{2} \sigma_z^{(k)} &+&
  \sum_{j=\mathrm{A},\mathrm{B}} (\epsilon\,\sigma_+^{(k)}\sigma_-^{(\mathrm{X})} + h.c.) \,.
\end{eqnarray}
(In Ref.\cite{ref:NakazatoPRA2004} the case of a different free Bohr frequency for the two-level system $\mathrm{X}$ is initially considered, but then the author focus on the homogeneous model with the three $\omega$'s all equal.)

The two terms of the Hamiltonian will be addressed as $H_0$ (the free part) and $H_\mathrm{AXB}$ (the term of interaction between $\mathrm{X}$ and the other two subsystems).

Now we want to consider the possibility of adding interaction terms between the subsystems $\mathrm{A}$ and $\mathrm{B}$. We consider the following:
\begin{eqnarray}
  H_\mathrm{AB} =
  \eta\,\ee^{\ii\phi}\,\sigma_+^{(A)}\sigma_-^{(\mathrm{B})} + h.c. \,,
\end{eqnarray}
which contains a direct interaction between such subsystems.

\subsection{Purification Scheme and Witness Quantities}

Following the typical scheme of purification previously mentioned, we assume that the system $\mathrm{X}$ is repeatedly measured and found in the same state $\Ket{\theta, \phi}_\mathrm{X}=\cos\theta\Ket{\uparrow}+\ee^{-\ii\phi}\sin\theta\Ket{\downarrow}$. The predictions about the extraction are given by diagonalizing
the operator
\begin{equation}
V(\tau) = _\mathrm{X}\Bra{\theta,\phi} U(\tau) \Ket{\theta,\phi}
_\mathrm{X} \,,
\end{equation}
which will be our reference model.

The eigenstate $\Ket{\lambda_1}$ corresponding to the largest eigenvalue (in modulus) of $V(\tau)$ is the one that will be extracted if the ancilla system $\mathrm{X}$ is repeatedly measured every $\tau$ and found in the same state. Since we want to extract an entangled state (preferably a maximally entangled state) we need to evaluate this feature. To this scope we can use the purity of the reduced density operator: ${\cal E}(\rho_{\mathrm{AB}})=2(1-P(\mathrm{tr}_\mathrm{A}\rho_\mathrm{AB}))$. We then introduce the parameter measuring the extracted entanglement as follows:
\begin{equation}
  \Upsilon = {\cal E}(|\lambda_1\rangle \langle \lambda_1 |)\,.
\end{equation}

The ratio between the largest ($\lambda_1$) and the second largest ($\lambda_2$) eigenvalue tells us the rapidity of the extraction process: the higher $|\lambda_1/\lambda_2|$, the smaller the number of steps required to extract the state. Therefore, here we introduce the efficiency as:
\begin{equation}
\Lambda = 1-\left|\lambda_2 /\lambda_1\right|^2\,.
\end{equation}
Of course, if the parameter $\Lambda$ is equal to $0$, there is no extraction of a single state. To have extraction we need this parameter to be nonzero. To have efficient extraction we need this ratio to be close to unity.

We also introduce the stability parameter,
\begin{equation}
  \Sigma = |\lambda_1|^2\,,
\end{equation}
which should approach $1$ to have an optimal extraction, otherwise the probability of success of the process will diminish at every step.

\section{Sensitivity to Undesired Couplings}\label{sec:stability}

\begin{figure}
\subfigure[]{\includegraphics[width=0.25\textwidth, angle=0]{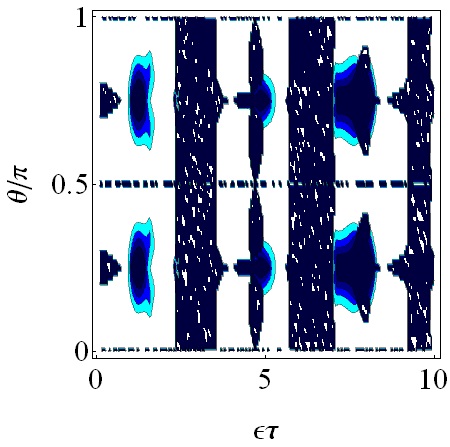}} \qquad%
\subfigure[]{\includegraphics[width=0.25\textwidth, angle=0]{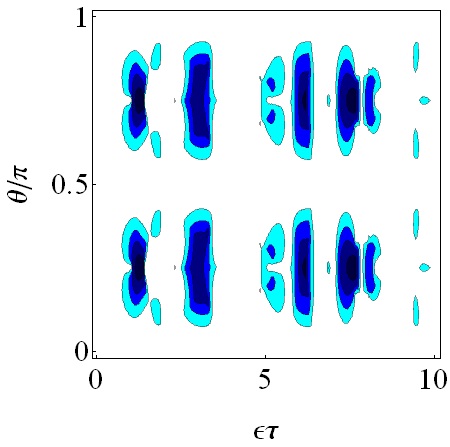}} \qquad%
\subfigure[]{\includegraphics[width=0.25\textwidth, angle=0]{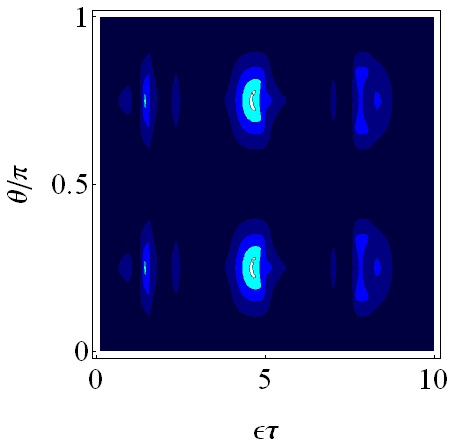}} \qquad%
\caption{(Color online). Entanglement $\Upsilon$ (a), Efficiency
$\Lambda$ (b) and Stability $\Sigma$ (c) of the extraction process
as functions of $\epsilon\tau$ and $\theta/\pi$, for
$\omega/\epsilon=2$, $\eta=0$. (All plotted quantities lie in the
range $[0,1]$, from white to dark blue.)} \label{fig:Standard}
\end{figure}

In Fig.\ref{fig:Standard} are shown the amount of entanglement of
the extracted state ($\Upsilon$), the efficiency ($\Lambda$) and
the stability ($\Sigma$) of the process of extraction in a case
with $\eta=0$, which is essentially the case analyzed in detail in
Ref.\cite{ref:NakazatoPRA2004}. It is clear that there are several
points (areas) where the entanglement of the extracted state is
very high (dark blue regions) and correspondingly the efficiency
and the stability are high too. This numerical result is in
perfect agreement with the theoretical analysis developed in
Ref.\cite{ref:NakazatoPRA2004}, where points of optimal extraction
of maximally entangled states have been found.

In the following subsections, we will consider the effects of the
perturbations previously described. In particular we will explore
the weak and the strong regime.

\subsection{Weak Coupling}

\begin{figure}
\subfigure[]{\includegraphics[width=0.25\textwidth, angle=0]{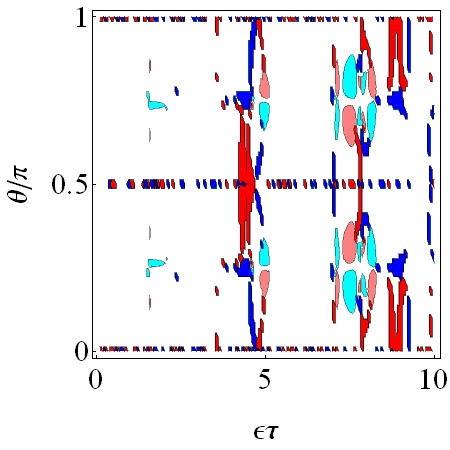}} \qquad%
\subfigure[]{\includegraphics[width=0.25\textwidth, angle=0]{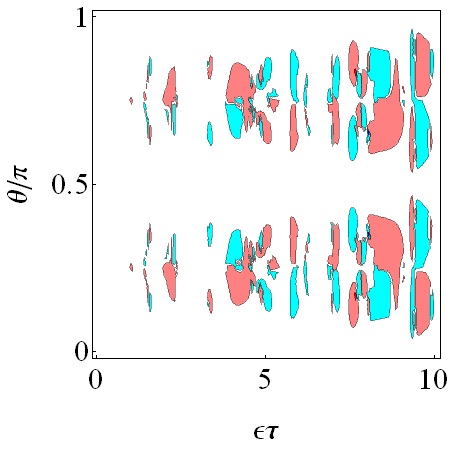}} \qquad%
\subfigure[]{\includegraphics[width=0.25\textwidth, angle=0]{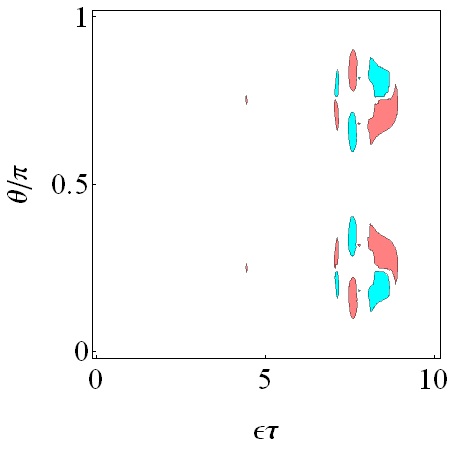}} \qquad%
\caption{(Color online). Entanglement $\Upsilon$ (a), Efficiency
$\Lambda$ (b) and Stability $\Sigma$ (c) of the extraction process
as functions of $\epsilon\tau$ and $\theta/\pi$, for
$\omega/\epsilon=2$, $\eta/\epsilon=0.01$ and $\phi=0$. (White
color corresponds to a discrepancy smaller than $0.01$; light blue
(red) means an increase (diminish) between $0.01$ and 0.1; dark
blue (red) means a discrepancy higher than $0.1$.)}
\label{fig:small1}
\end{figure}

\begin{figure}
\subfigure[]{\includegraphics[width=0.25\textwidth, angle=0]{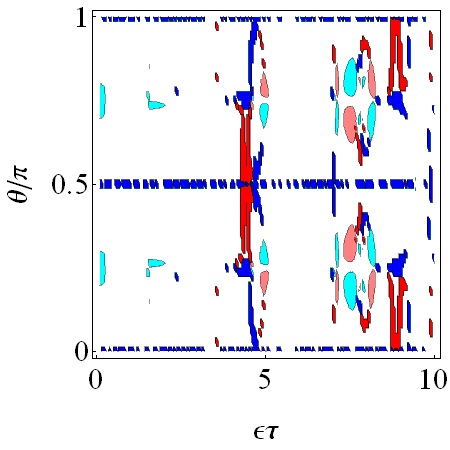}} \qquad%
\subfigure[]{\includegraphics[width=0.25\textwidth, angle=0]{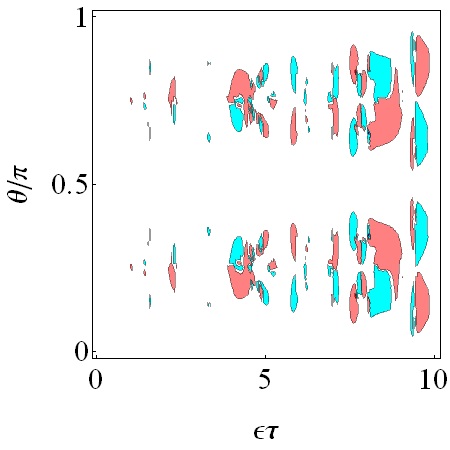}} \qquad%
\subfigure[]{\includegraphics[width=0.25\textwidth, angle=0]{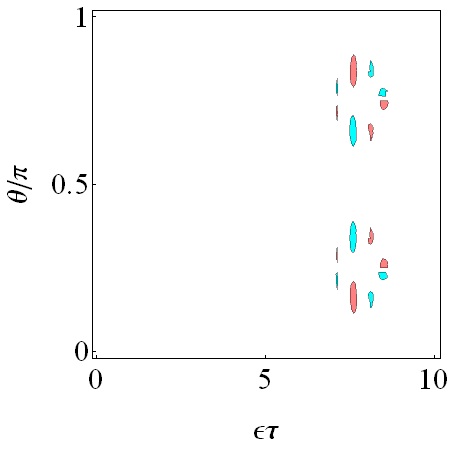}} \qquad%
\caption{(Color online). Entanglement $\Upsilon$ (a), Efficiency
$\Lambda$ (b) and Stability $\Sigma$ (c) of the extraction process
as functions of $\epsilon\tau$ and $\theta/\pi$, for
$\omega/\epsilon=2$, $\eta/\epsilon=0.01$ and $\phi=\pi/4$. (White
color corresponds to a discrepancy smaller than $0.01$; light blue
(red) means an increase (diminish) between $0.01$ and 0.1; dark
blue (red) means a discrepancy higher than $0.1$.)}
\label{fig:small2}
\end{figure}

\begin{figure}
\includegraphics[width=0.25\textwidth, angle=0]{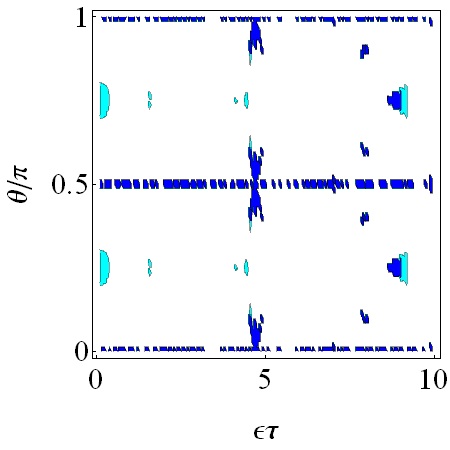}
\caption{(Color online). Entanglement $\Upsilon$ of the extraction
process as functions of $\epsilon\tau$ and $\theta/\pi$, for
$\omega/\epsilon=2$, $\eta/\epsilon=0.01$ and $\phi=\pi/2$.
Graphics for Efficiency and Stability are omitted, since they are
white rectangles, meaning that there are no significant
discrepancies with the $\eta=0$ case. (White color corresponds to
a discrepancy smaller than $0.01$; light blue (red) means an
increase (diminish) between $0.01$ and 0.1; dark blue (red) means
a discrepancy higher than $0.1$.)} \label{fig:small3}
\end{figure}

An additional coupling $H_\mathrm{AB}$ can alter the results of
the extraction process, even for small values of $\eta$. In
Figs.~\ref{fig:small1},\ref{fig:small2},\ref{fig:small3} are shown
the discrepancies of the amount of entanglement of the extracted
state, of the efficiency and of the stability of the process, as
functions of $\theta$ and $\tau$, for particular values of $\eta$,
$\phi$ and $\epsilon$, with respect to the $\eta=0$ case. Blue
zones indicate improvements (higher entanglement, efficiency or
stability, depending on the case), red zones indicate worsenings,
while white areas indicate zero difference.

In Fig.~\ref{fig:small1} it is considered the case
$\eta/\epsilon=0.01$ and $\phi=0$. It is well visible that there
are wide zones white colored, meaning that there are a lot of
cases (values of $\theta$ and $\tau$) where the process results
insensitive to the presence of the inner coupling. Anyway,
differences are present for entanglement, efficiency and
stability. In particular, as for the efficiency, it is well
visible that for small values of the variable $\tau$ (say for
$\epsilon\tau < 6$) there are no changes of efficiency due to
$H_\mathrm{AB}$.

Fig.~\ref{fig:small2} describes almost the same situation of
Fig.~\ref{fig:small1} except for the phase, which now assumes the
value $\phi=\pi/4$. This modified value of $\phi$ produces visible
differences, which are very significant in the efficiency, where
the white zones are wider than in the $\phi=0$ case.

In Fig.~\ref{fig:small3} it is shown only the variation of the
extracted entanglement for the $\phi=\pi/2$ case, because
efficiency and stability graphics are white rectangles, meaning
that there are no differences with the $\eta=0$ case. As for the
entanglement, in this case there are only blue zones, suggesting
that for small $\eta$ and $\phi=\pi/2$ there are only possible
improvements of the extracted entanglement, never a worsening.

\subsection{Strong Coupling}

\begin{figure}
\subfigure[]{\includegraphics[width=0.25\textwidth, angle=0]{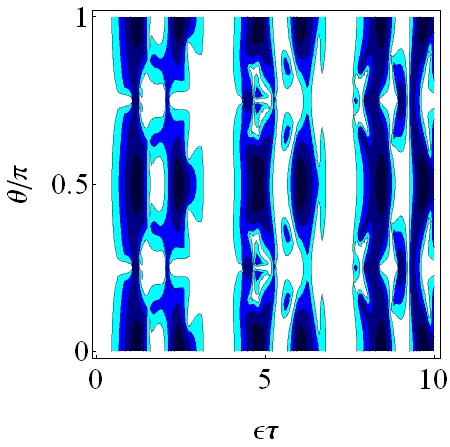}} \qquad 
\subfigure[]{\includegraphics[width=0.25\textwidth, angle=0]{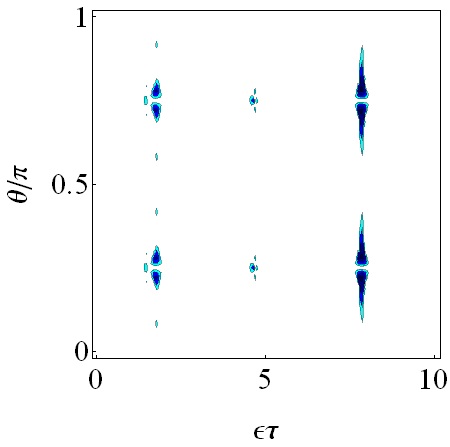}} \qquad %
\subfigure[]{\includegraphics[width=0.25\textwidth, angle=0]{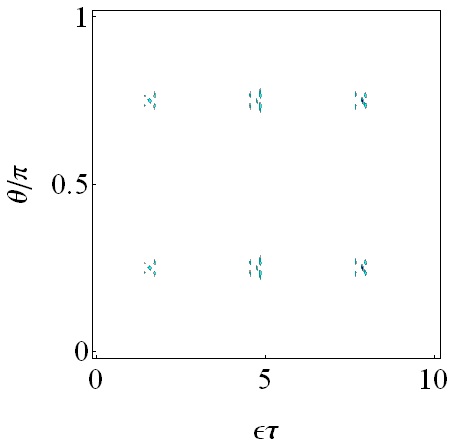}} 
\caption{(Color online). Efficiency of extraction $\Lambda$ as a
function of $\epsilon\tau$ and $\theta/\pi$, with
$\eta/\epsilon=1$ (a), $\eta/\epsilon=5$ (b), $\eta/\epsilon=20$
(c). Here $\omega/\epsilon=2$ and $\phi=\pi/2$ in all plots. The
plot corresponding to higher values of $|\eta/\epsilon|$ are not
reported here, since they simply are white rectangles. (All
plotted quantities lie in the range $[0,1]$; a darker blue
corresponds to a higher value.)} \label{fig:BigPi1}
\end{figure}

\begin{figure}
\subfigure[]{\includegraphics[width=0.25\textwidth, angle=0]{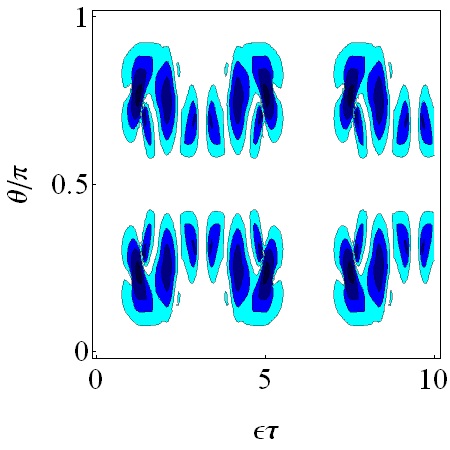}} \qquad 
\subfigure[]{\includegraphics[width=0.25\textwidth, angle=0]{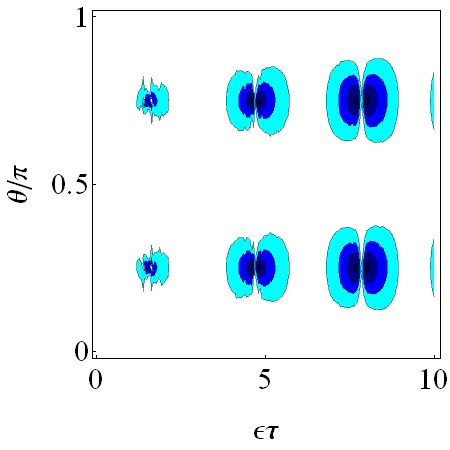}} \qquad %
\subfigure[]{\includegraphics[width=0.25\textwidth, angle=0]{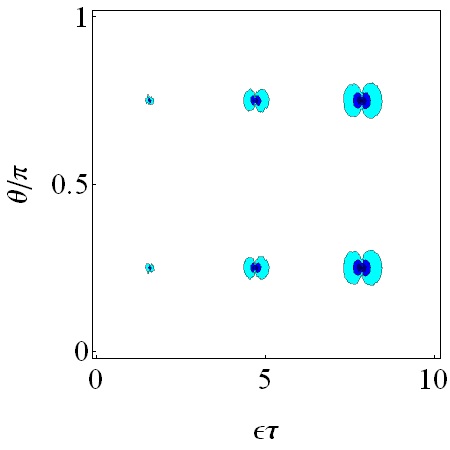}} 
\caption{(Color online). Efficiency of extraction $\Lambda$ as a
function of $\epsilon\tau$ and $\theta/\pi$, with
$\eta/\epsilon=1$ (a), $\eta/\epsilon=25$ (b), $\eta/\epsilon=100$
(c). Here $\omega/\epsilon=2$ and $\phi=0$ in all plots. The plot
corresponding to higher values of $|\eta/\epsilon|$ are not
reported here, since they simply are white rectangles. (All
plotted quantities lie in the range $[0,1]$; a darker blue
corresponds to a higher value.)} \label{fig:BigPi2}
\end{figure}

It is pretty intuitive that in the strong coupling limit
($\eta\gg\epsilon$) the disturbance from $H_\mathrm{AB}$ is more
significant, and the patterns for entanglement, efficiency and
stability are very different from this one case for $\eta=0$. In
particular, we have observed that for increasing values of $\eta$
the efficiency of the extraction process approaches zero almost
everywhere, i.e., for every values of $\theta$ and $\tau$.
Figs.~\ref{fig:BigPi1} and \ref{fig:BigPi2} show this phenomenon
in a clear way. Moreover, also in this case a sensitivity to the
phase $\phi$ is very well visible.

In Fig.~\ref{fig:BigPi1} the $\phi=\pi/2$ case is considered for
increasing values of $\eta$: $\eta/\epsilon=1$, $\eta/\epsilon=5$
and $\eta/\epsilon=20$, which show that the higher $\eta$ the
smaller the efficiency (whiter picture). We have made plots
corresponding to values of $\eta/\epsilon$ higher than $20$, but
they are not reported here, since they are simply white
rectangles. Fig.~\ref{fig:BigPi2} shows the behaviour of the
efficiency for the $\phi=0$ case. The effect of a diminishing
efficiency is still present, but this time we need higher values
of $\eta$ to obtain something comparable to what happens for
$\phi=\pi/2$.

\section{Theoretical analysis}\label{sec:qze}

A complete theoretical explanation of the behaviour of the witness quantities in the different regimes (different values of $\eta$ and $\phi$) would require detailed mathematical analysis of the $V(\tau)$ operator. Though in our case it is a $4\times 4$ matrix, its entries are very long complicated expressions, and the complete diagonalization is not easy, nor the results are readable. Nevertheless, it is possible to reach some conclusions through some qualitative arguments.

First of all, let us consider the structure of the Hamiltonian:
\begin{equation}
H = \left(
\begin{array}{cccccccc}
3\omega & 0 & 0 & 0 & 0 & 0 & 0 & 0 \\
0 & 2\omega & \eta\,\ee^{\ii\phi} & \epsilon & 0 & 0 & 0 & 0 \\
0 & \eta\,\ee^{-\ii\phi} & 2\omega & \epsilon & 0 & 0 & 0 & 0 \\
0 & \epsilon & \epsilon & 2\omega & 0 & 0 & 0 & 0 \\
0 & 0 & 0 & 0 & \omega & \eta\,\ee^{\ii\phi} & \epsilon & 0 \\
0 & 0 & 0 & 0 & \eta\,\ee^{-\ii\phi} & \omega & \epsilon & 0 \\
0 & 0 & 0 & 0 & \epsilon & \epsilon & \omega & 0 \\
0 & 0 & 0 & 0 & 0 & 0 & 0 & 0 \\
\end{array}\right)\,,
\end{equation}
which is given with respect to the following basis:
$\Ket{\uparrow\uparrow}\Ket{\uparrow}_\mathrm{X}$,
$\Ket{\uparrow\downarrow}\Ket{\uparrow}_\mathrm{X}$,
$\Ket{\downarrow\uparrow}\Ket{\uparrow}_\mathrm{X}$,
$\Ket{\uparrow\uparrow}\Ket{\downarrow}_\mathrm{X}$,
$\Ket{\uparrow\downarrow}\Ket{\downarrow}_\mathrm{X}$,
$\Ket{\downarrow\uparrow}\Ket{\downarrow}_\mathrm{X}$,
$\Ket{\downarrow\downarrow}\Ket{\uparrow}_\mathrm{X}$,
$\Ket{\downarrow\downarrow}\Ket{\downarrow}_\mathrm{X}$.

In the weak coupling limit, the eigenstates of the Hamiltonian can
be found through a perturbation treatment in the parameter
$\eta/\epsilon$, where the unperturbed Hamiltonian is $H_0 +
H_\mathrm{AXB}$ and the perturbation is $H_\mathrm{AB}$. The
unperturbed eigenvalues and eigenstates are: $3\omega$,
$2\omega+\epsilon\sqrt{2}$, $2\omega-\epsilon\sqrt{2}$, $2\omega$,
$\omega+\epsilon\sqrt{2}$, $\omega-\epsilon\sqrt{2}$, $\omega$,
$0$, and $\Ket{\uparrow\uparrow}\Ket{\uparrow}_\mathrm{X}$,
$2^{-1/2}(\Ket{\uparrow\uparrow}\Ket{\downarrow}_\mathrm{X}+\Ket{\Psi^S}\Ket{\uparrow}_\mathrm{X})$,
$2^{-1/2}(\Ket{\uparrow\uparrow}\Ket{\downarrow}_\mathrm{X}-\Ket{\Psi^S}\Ket{\uparrow}_\mathrm{X})$,
$\Ket{\Psi^A}\Ket{\uparrow}_\mathrm{X}$,
$2^{-1/2}(\Ket{\downarrow\downarrow}\Ket{\uparrow}_\mathrm{X}+\Ket{\Psi^S}\Ket{\downarrow}_\mathrm{X})$,
$2^{-1/2}(\Ket{\downarrow\downarrow}\Ket{\uparrow}_\mathrm{X}-\Ket{\Psi^S}\Ket{\downarrow}_\mathrm{X})$,
$\Ket{\Psi^A}\Ket{\downarrow}_\mathrm{X}$,
$\Ket{\downarrow\downarrow}\Ket{\downarrow}_\mathrm{X}$,
respectively, with
$\Ket{\Psi^S}=2^{-1/2}(\Ket{\downarrow\uparrow}+\Ket{\uparrow\downarrow})$
and
$\Ket{\Psi^A}=2^{-1/2}(\Ket{\downarrow\uparrow}-\Ket{\uparrow\downarrow})$.
The first order corrections to the eigenstates are of the order
$\eta/\epsilon$, and the corrections to the eigenvalues are: $0$,
$(\eta/2)\cos\phi$, $(\eta/2)\cos\phi$, $-\eta\cos\phi$,
$(\eta/2)\cos\phi$, $(\eta/2)\cos\phi$, $-\eta\cos\phi$, $0$. This
means that for $\phi=\pi/2$ there are no corrections to the
eigenvalues, which makes the evolution operator and the $V(\tau)$
operator closer to the $\eta=0$ counterparts, somehow supporting
the numerical result that for $\phi=\pi/2$ both efficiency and
stability do not exhibit significant discrepancies with respect to
the $\eta=0$ case.

The strong coupling limit can again be analyzed through a perturbation treatment with respect to $\epsilon/\eta$. The eigenvaues and eigenstates of the unperturbed Hamiltonian $H_0+H_\mathrm{AB}$ are: $3\omega$, $2\omega+\eta$, $2\omega-\eta$, $2\omega$, $\omega+\eta$, $\omega-\eta$, $\omega$, $0$ and
$\Ket{\uparrow\uparrow}\Ket{\uparrow}_\mathrm{X}$,
$2^{-1/2}(\Ket{\uparrow\downarrow}+\ee^{-\ii\phi}\Ket{\downarrow\uparrow})\Ket{\uparrow}_\mathrm{X}$,
$2^{-1/2}(\Ket{\uparrow\downarrow}-\ee^{-\ii\phi}\Ket{\downarrow\uparrow})\Ket{\uparrow}_\mathrm{X}$,
$\Ket{\uparrow\uparrow}\Ket{\downarrow}_\mathrm{X}$,
$\Ket{\downarrow\downarrow}\Ket{\uparrow}_\mathrm{X}$,
$2^{-1/2}(\Ket{\uparrow\downarrow}+\ee^{-\ii\phi}\Ket{\downarrow\uparrow})\Ket{\downarrow}_\mathrm{X}$,
$2^{-1/2}(\Ket{\uparrow\downarrow}-\ee^{-\ii\phi}\Ket{\downarrow\uparrow})\Ket{\downarrow}_\mathrm{X}$,
$\Ket{\downarrow\downarrow}\Ket{\downarrow}_\mathrm{X}$, respectively. The corrections to the eigenstates are of the order $\epsilon/\eta$, becoming more and more negligible when $\eta$ increases; the corrections to the eigenvalues are all zero. Therefore, on the one hand, it is easy to understand that for very large $\eta$ the influence of $H_\mathrm{AXB}$ on the dynamics becomes negligible, meaning that the subsystems $\mathrm{AB}$ and $\mathrm{X}$ can be considered as decoupled, then jeopardizing the extraction process. (This occurrence can be seen as a generalized QZE~\cite{ref:Schulman,ref:Panov}, in the sense of a Hilbert space partitioning~\cite{ref:FacchiPasc2002,ref:Militello2001,ref:Militello2011}.) On the other hand, in this case there is no easy and direct explanation of the phase effect consisting in an acceleration of the efficiency diminishing when $\phi=\pi/2$. It can be understood in terms of a complete diagonalization of $V(\tau)$, which, however, is beyond the scope of the present work.

\section{Conclusions}\label{sec:conclusions}

In this paper we have reconsidered the purification scheme introduced in Ref.~\cite{ref:NakazatoPRL2003}, in particular analyzing the possibility of taking into account additional interactions to a prefixed scheme. We focused on the special regimes of weak and strong coupling.

The additional interaction that we have considered, seemingly, should be helpful for the establishment of an entaglement between the subsystems $\mathrm{A}$ and $\mathrm{B}$. Neverheless, depending on the situation, it can be helpful or harmful to the extraction process. The numerical predictions are partly supported by a theoretical semi-quantitative analysis valid in the weak and strong coupling limit. In this second case, a dramatic dimishing of the efficiency is predicted, and its connection with a generalized quantum Zeno effect (in the sense of an interaction-induced partitioning of the relevant Hilbert space) is demonstrated.

It is worth recalling that originally the QZE has been introduced as the possibility of hindering a natural decay (of an atom) through repeated pulsed measurements. Subsequently, the possibility of a dynamical inhibition through strong decays or
strong additional couplings has been proven, leading to the notion of a generalized QZE based on Hilbert space partitioning. Now, in this paper, we have considered the effects of an additional interaction that can somehow neutralize the effects of repeated pulsed measurements on a system. This fact clearly shows how rich is the panorama of all possible interplays between interactions and iterated measurements, beyond the standard formulation of the QZE.





\end{document}